\documentclass[pra,prlaprintnumbers,twocolumn,superscriptaddress]{revtex4}
\usepackage{amsmath}
\usepackage{dcolumn}
\usepackage{bm}
\usepackage{epsfig}
\usepackage{epstopdf}
\usepackage{graphicx}
\usepackage{amsfonts}
\usepackage{amssymb}
\usepackage{mathrsfs}
\usepackage{color}
\usepackage{multirow}
\usepackage{appendix}

\newcommand{\bra}[1]{\langle#1\rvert} 
\newcommand{\ket}[1]{\lvert#1\rangle} 

\newcommand{\ZZU}{Quantum Information Institute, School of Physics and Laboratory of Zhongyuan Light,
Zhengzhou University, Zhengzhou 450001, China}

\newcommand{\HAS}{Institute of Quantum Materials and Physics, Henan Academy of Sciences, Zhengzhou 450046, China}

\begin{document}
\title{Experimental Demonstration of a Brachistochrone Nonadiabatic Holonomic Quantum-Gate Scheme in a Trapped Ion}

\author{Xi Wang}
\altaffiliation{Co-first authors have the same contributions}
\affiliation{\ZZU}

\author{Hui Ren}
\altaffiliation{Co-first authors have the same contributions}
\affiliation{\ZZU}

\author{L.-N. Sun}
\affiliation{\ZZU}

\author{K.-F. Cui}
\email{cuikaifeng@zzu.edu.cn}
\affiliation{\ZZU}

\author{J.-T. Bu}
\affiliation{\ZZU}

\author{S.-L. Su}
\affiliation{\ZZU}
\affiliation{\HAS}

\author{L.-L. Yan}
\email{llyan@zzu.edu.cn}
\affiliation{\ZZU}
\affiliation{\HAS}

\author{G. Chen}
\affiliation{\ZZU}

\begin{abstract}
Nonadiabatic holonomic quantum computation (NHQC) offers intrinsic resilience to certain control imperfections. However, conventional nonadiabatic holonomic protocols are constrained by the fixed-pulse-area condition, which limits flexibility and prolongs duration of small-angle gates. Here we experimentally demonstrate a universal brachistochrone nonadiabatic holonomic quantum gate scheme in a trapped $^{40}\mathrm{Ca}^{+}$ ion, and realized the construction of $\sqrt{X}$ gate under the conventional NHQC, brachistochrone NHQC (BNHQC) and composite BNHQC (CBNHQC) protocols. By characterizing the performance of gate performance in the presence of dissipation, Rabi-frequency errors and detuning errors, we show that BNHQC and CBNHQC outperform conventional NHQC, and BNHQC can offer a favorable balance between operation speed and robustness. It further shows that keeping high fidelity and strong robustness need decrease the accumulated population of excited state in the evolution process. These results highlight nonadiabatic holonomic computation as a practical route toward fast and robust quantum gates in trapped-ion platforms.
\end{abstract}

\maketitle
\textit{Introduction}$-$Universal quantum computation not only requires high-fidelity quantum gates, also requires them to be robust against both systematic control errors~\cite{ref1,ref2,ref3} and environmental decoherence~\cite{ref4,ref5,ref6}. In this context, holonomic quantum computation (HQC),  benefiting from the global features of evolution path, has emerged as a promising approach to robust quantum control~\cite{ref7,ref8,ref9,ref10,ref11,ref12}. However, HQC is originally formulated in the adiabatic regime and results in a long gate duration. To overcome this limitation, nonadiabatic holonomic quantum computation (NHQC) is proposed~\cite{ref13,ref14,ref15,ref16}. By exploiting non-Abelian geometric phases, NHQC further shows the speed advantage of nonadiabatic driving, and has been experimentally demonstrated in different systems, such as superconducting circuits~\cite{ref17,ref18,ref19,ref20,ref21,ref22,ref23,ref24,ref25}, nuclear magnetic resonance~\cite{ref23,ref24}, semiconductor quantum dots~\cite{ref25,ref26}, nitrogen-vacancy centers in diamond~\cite{ref27,ref28,ref29,ref30,ref31}, and trapped ions~\cite{ref32,ref33}.

Despite these advances, conventional NHQC still suffers from the fixed-pulse-area constraint regardless of the target rotation angle~\cite{ref15,ref34}. As a result, the gate duration remains unchanged even for small-angle rotations, unnecessarily increasing the qubit's exposure to environmental noise. To overcome this limitation, the framework of brachistochrone nonadiabatic holonomic quantum computation (BNHQC)~\cite{ref35,ref36,ref37,ref38,ref50,ref39,ref40} is proposed by combining NHQC with time-optimal control (TOC) to further decrease the duration of gate.

In this work, we experimentally construct a universal quantum gate by the NHQC and BNHQC in a single trapped $^{40}\mathrm{Ca}^{+}$ ion system, and demonstrate the advantage of BNHQC. 
By comparing the speed and fidelity of conventional NHQC, BNHQC and composite BNHQC (CBNHQC)~\cite{ref45,ref46}, we systematically characterize their robustness against decoherence and control errors of Rabi-frequency and detuning. Our results show that BNHQC and CBNHQC outperform conventional NHQC in both the fidelity and robustness. CBNHQC provides stronger suppression of systematic errors via composite pulse symmetry but with a longer duration, while BNHQC reduces decoherence by minimizing the evolution time. In practice, BNHQC achieves an excellent balance between operation speed and noise resilience. 


\begin{figure*}[htpb]
\includegraphics[width=17.0cm,height=4.5 cm]{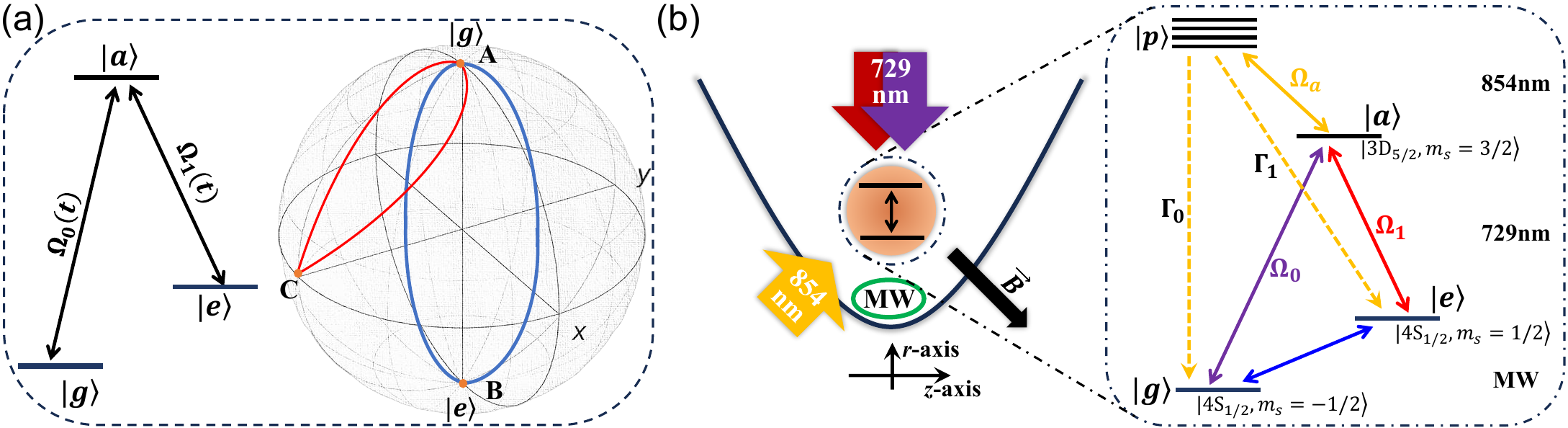}
\includegraphics[width=17.0cm,height=4.8 cm]{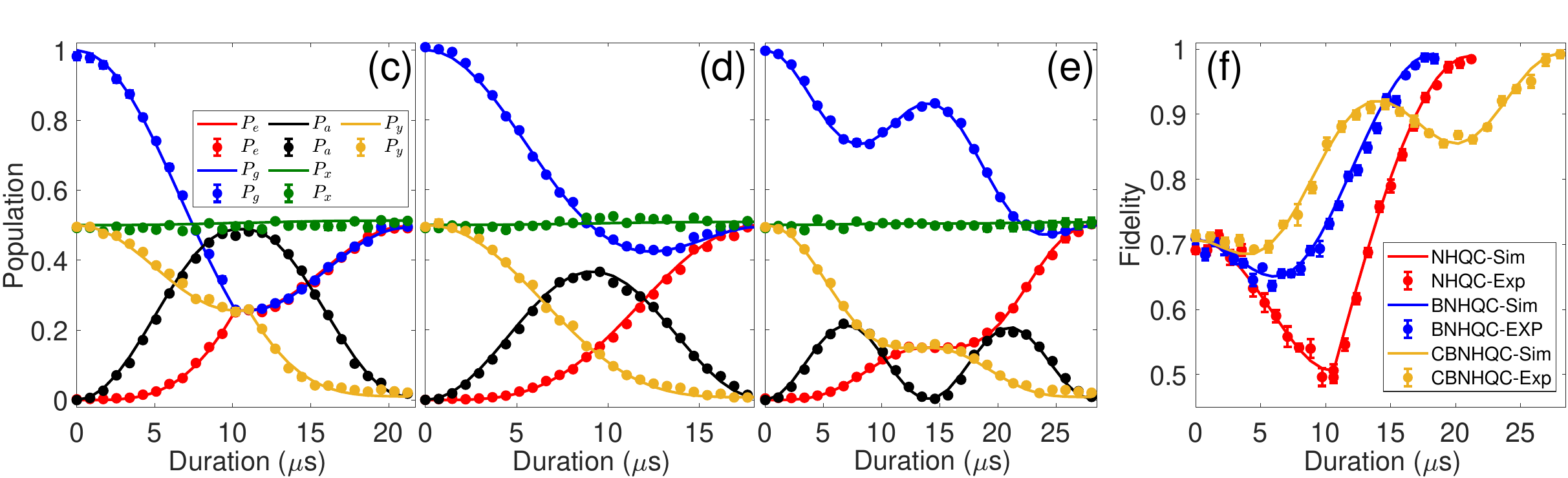}
\caption{Experimental demonstration of universal quantum gate in a trapped ion. (a) Schematic for realizing a universal single quantum gate: (left) the energy-level and (right) the trajectories of realizing a conventional dynamical gate (blue) and a geometric gate based on NHQC (red) on the Bloch sphere. (b) Set-up of realizing a universal single quantum gate in a single trapped $^{40}\mathrm{Ca}^{+}$ ion. (c)-(e) Experimentally measured population evolution for the $\sqrt{X}$ gate implemented by NHQC, BNHQC, and CBNHQC, respectively, 
(f) The evolution of fidelity with respect to the ideal target state. Here the initial state is $\ket{g}$, the Rabi frequency $\Omega/2\pi=47.1$~kHz, the effective decay rate of auxiliary state $\kappa=5$~kHz and the error bars indicating the statistical standard deviation of the experimental data are obtained by 20000 measurements for each data point.}
\label{fig1}
\end{figure*}

\textit{Control protocols}$-$Before specifying the experimental demonstration, we first elucidate the theory of constructing a universal single-qubit gate using these control protocols. Considering a $\Lambda$-type three-level system, as shown in Fig.~\ref{fig1}(a), a qubit is encoded in states $\ket{g}$ and $\ket{e}$, and an auxiliary state $\ket{a}$ is coupled to the qubit by two driving lasers~\cite{ref13,ref14} with the interaction Hamiltonian as
\begin{equation}
H = \frac{\Omega(t)}{2}e^{i\phi_{1}(t)}  [ e^{i\phi(t) } \sin\frac{\theta}{2} \ket{g} + \cos \frac{\theta}{2}\ket{e}]  \bra{a} + \text{H.C.},
\end{equation}
where $\Omega(t)=\sqrt{\Omega_0^2(t)+\Omega_1^2(t)}$ and $\phi(t)=\phi_0(t)-\phi_1(t)$ with the Rabi frequencies and phases of lasers corresponding to $\Omega_{0,1}(t)$ and $\phi_{0,1}(t)$.

For the NHQC, the pulse amplitudes of two lasers keep constant ($\Omega(t)\equiv\Omega$) for the total duration $\tau_N=2\pi/\Omega$, while the phases are controlled in two equal halves: During the first half, the phases are set as $(\phi,0)$, and jump to $(\phi+\gamma,\gamma)$ at the midpoint for the second half, which thus produces an unsmooth trajectory (see Fig.~\ref{fig1}(a)). After eliminating the auxiliary level, the effective evolution of qubit is obtained as $U(\theta ,\phi,\gamma)= \exp(-i\frac{\gamma}{2}\mathbf{n} \cdot\bm{\sigma})$ with the Bloch vector $\bm{n} =(\sin \theta \cos \phi, \sin \theta \sin \phi, \cos \theta)$ and Pauli vector $\bm{\sigma} = (\sigma_x, \sigma_y, \sigma_z)$. By choosing different parameter sets of $\{\theta$, $\phi$, $\gamma\}$, we can realize a universal geometric gates of single-qubit. For example, we can set $ \theta  = \pi /2,\phi  = 0,\gamma  = \pi /2 $ to realize the $\sqrt{X}$ gate with evolution operator $ {U}_{\sqrt{X}} = ({I - i{\sigma }_{x}}) /\sqrt{2} $, and set $ \theta  = 0,\phi  = 0,\gamma  = \pi /4 $ to implement the $T$ gate with evolution operator $ {U}_{T} = \ket{e}\bra{e}-e^{-i\pi/4}\ket{g}\bra{g} $. 
In this work, we use the $\sqrt{X}$ gate as a representative benchmark and also set $\Omega(t)$ as a constant for the following two protocols.

\begin{figure*}[htpb]
\includegraphics[width=17.0cm,height=6.4cm]{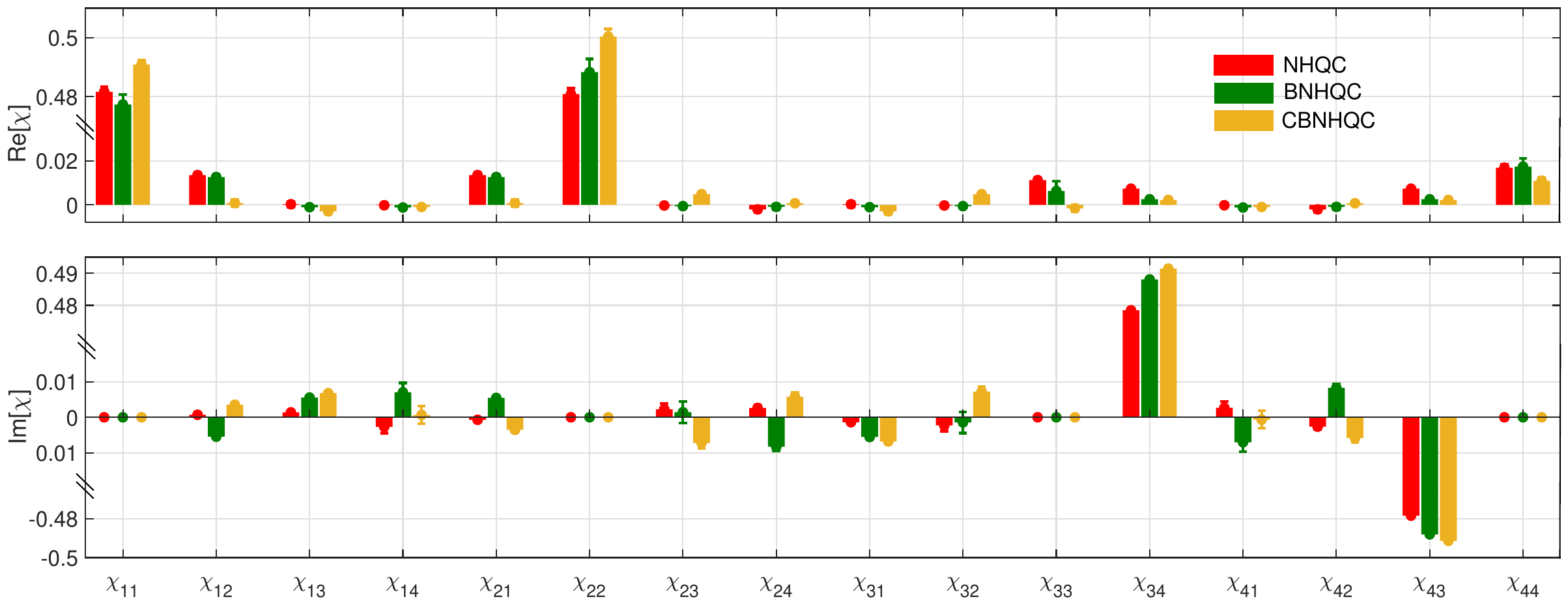}
\caption{The experimentally measured histogram charts for the real and imaginary parts of process matrix $\chi$ of the $\sqrt{X}$ gate under different control protocols, where the data around 0 and 0.5 are enlarged to display the difference among different gate schemes and
distinguish small errors of the experimental data. Here the parameters selected same as Fig.\ref{fig1}.}
\label{fig:matrix}
\end{figure*}

To settle the fixed-pulse-area constraint of NHQC, BNHQC is proposed by replacing the abrupt phase jump with a continuous time-optimal phase modulation, which can be obtained by
\begin{equation}
\Omega_{0}(t) = \Omega e^{i(\phi + \phi_1 )}\sin\frac{\theta}{2},\ \Omega_{1}(t) = \Omega e^{i\phi_{1}}\cos\frac{\theta}{2}. 
\end{equation}
By solving the quantum brachistochrone equation (QBE)~\cite{ref41,ref42,ref43,ref44}, we can obtain the time-dependent phase and time-optimal duration corresponding to 
\begin{equation}
\phi_{1}(t)=2(\pi-\gamma)t/\tau_{B}, \ \tau_B = 2\sqrt{\pi^2-(\pi-\gamma)^2}/\Omega,
\end{equation}
which makes the duration largely decreased for the small-angle rotations. For instance, BNHQC realizes the $\sqrt{X}$ gate with $ \tau_{B} =\sqrt{3}\pi/\Omega$, shorter than $\tau_{N}=2\pi/\Omega$.

To further enhance robustness against control errors, we combine BNHQC by a composite-pulse construction to produce a CBNHQC. A simple realization of CBNHQC is using two segments, each of which is corresponding to a BNHQC with a rotation angle $\gamma/2$ and constant Rabi frequency. The phase of laser is controlled as
\begin{equation}
\left\lbrace 
\begin{aligned}
&\phi_1(t)=2(\pi-\gamma/2)t/\tau_C,\quad t\in [0,\tau_C/2] \\
&\phi_1(t)=\pi+2(\pi-\gamma/2)t/\tau_C,\quad t\in [\tau_C/2,\tau_C]
\end{aligned}
\right.
\end{equation} 
with the duration of gate as 
\begin{equation}
\tau_C=4\sqrt{\pi^2-(\pi-\gamma/2)^2}/\Omega,
\end{equation}
which will give a longer duration than the NHQC and BNHQC, such as $ \tau_{C} = \sqrt{7}\pi/\Omega $ for the $\sqrt{X}$ gate. However, CBNHQC can give a higher fidelity due to the advantage of suppressing systematic control errors. In the following, we will experimentally compare these three protocols based on the demonstration of $\sqrt{X}$ gate.

\textit{Experimental system}$-$We perform the experiment in a five-segment linear ion trap fabricated by CIQTEK. As shown in Fig.~\ref{fig1}(b), a single $^{40}\text{Ca}^{+}$ ion is confined at the trap center by a radio-frequency field with $\omega_{\rm RF}/2\pi = 43.83~\mathrm{MHz}$. Under a driving power of about 4~W, the secular frequencies are $(\omega_x, \omega_y, \omega_z) = 2\pi \times (0.36, 1.89, 1.85)~\mathrm{MHz}$, where the weakest confinement direction defines the trap-$z$ axis. The qubit is encoded in the Zeeman sub-level of ground state as $\ket{g} \equiv \ket{4\mathrm{S}_{1/2}, m_s=-1/2}$ and $\ket{e} \equiv \ket{4\mathrm{S}_{1/2}, m_s=1/2}$. A magnetic field of 9.485~G is applied in the horizontal plane at an angle of $45^\circ$ to the $z$-axis, giving an energy splitting of $\omega_s/2\pi = 26.55~\mathrm{MHz}$ between $\ket{g}$ and $\ket{e}$. The auxiliary state is chosen as $\ket{a} \equiv \ket{3\mathrm{D}_{5/2}, m_s=+3/2}$ to form the $\Lambda$ system. A narrow-linewidth bichromatic 729~nm laser, propagating perpendicular to the trap $z$-axis and modulated by a dual-frequency signal, couples $\ket{g}$ and $\ket{e}$ to $\ket{a}$, respectively. 

The ion is initialized in either $\ket{g}$ or $\ket{e}$ by combining the optical pumping with a 397~nm circularly polarized laser and repumping with a series of 729~nm $\pi$ pulses and 854~nm laser. The state-preparation fidelity exceeds over 99.5\%. A radio-frequency coil, mounted opposite to the 729~nm laser, produces a microwave field at frequency $\omega_m$ to drive transitions between $\ket{e}$ and $\ket{g}$. By adjusting the microwave (MW) phase, we implement arbitrary single-qubit rotations and prepare the input states needed for the  tomography~\cite{ref47,ref48,ref49} of state and process. 
For detection, two 397~nm beams for precooling and Doppler cooling, together with 866~nm and 854~nm repumping beams, are applied perpendicular to the magnetic field. When the 397~nm and 866~nm lasers are tuned close to resonance, the ion in the $\ket{4\mathrm{S}_{1/2}}$ manifold emits fluorescence at about 60~kcps. In contrast, the auxiliary state $\ket{a}$ has a lifetime of about 1.2~s, and its fluorescence is below 1~kcps, so it serves as a dark state. To measure the population of a selected state, we apply a series of 729~nm $\pi$ pulses to shelve that state to the $\ket{4\mathrm{D}_{5/2}}$ manifold and then count the dark-state signal. 


As shown in Fig.~\ref{fig1}(b), an 854~nm laser couples $\ket{3\mathrm{D}_{5/2}}$ and $\ket{4\mathrm{P}_{3/2}}$ and provides the controlled dissipation of the auxiliary state. This laser addresses several transitions between $\ket{3\mathrm{D}_{5/2}, m_J=+3/2}$ and $\ket{4\mathrm{P}_{3/2}, m_J=\pm 1/2,\pm3/2}$. After adiabatic elimination of excited states, two effective decay channels are formed as $\ket{a}\rightarrow\ket{g}$ and $\ket{a}\rightarrow\ket{e}$ with the branching ratio $\kappa_g/\kappa_e = 3/22$. 
To perform complete quantum state tomography, we also measure the projections onto the basis $\ket{g}_x = (\ket{e} - \ket{g})/\sqrt{2}$ and $\ket{g}_y = (\ket{e} - i \ket{g})/\sqrt{2}$, which are realized by applying a $\pi/2$ microwave pulse before detection. 

\textit{Results}$-$To demonstrate the advantage of BNHQC, we first track the state evolution of the $\sqrt{X}$ gate for NHQC, BNHQC, and CBNHQC by the state tomography in Fig.~\ref{fig1}(c)-(e), where the density matrix of system is reconstructed by $\rho=(I+\bm{s}\cdot\bm{\sigma})/2$, where the Stokes parameter vector $\bm{s}=[s_x, s_y, s_z]$ with $s_x=s_0-2P_x$, $s_y=1-2P_y$, $s_z=1-2P_g$ and $s_0=P_g+P_e$. For an ideal $\sqrt{X}$ gate, the input state $\ket{g}$ should evolve to $(\ket{g}-i\ket{e})/\sqrt{2}$. From the evolution of state fidelity, as shown in Fig.~\ref{fig1}(f), it clearly demonstrates that the BNHQC has the fastest speed of constructing $\sqrt{X}$ gate. The final fidelities (defined as $F=\text{Tr}[\sqrt{\sqrt{\rho_{\rm exp}}\rho_{\rm id}\sqrt{\rho_{\rm exp}}}]$) are 98.5(4)\%, 98.6(7)\%, and 99.2(6)\% for NHQC, BNHQC, and CBNHQC, respectively, consistent with their theoretical prediction values 99.0\%, 99.3\% and 99.4\%.  In our experiment, the largest experimental errors arise from the state preparation and measurement errors, which normally below 0.5\% in our system.

\begin{figure}[t]
\includegraphics[width=8.3cm,height=9.3cm]{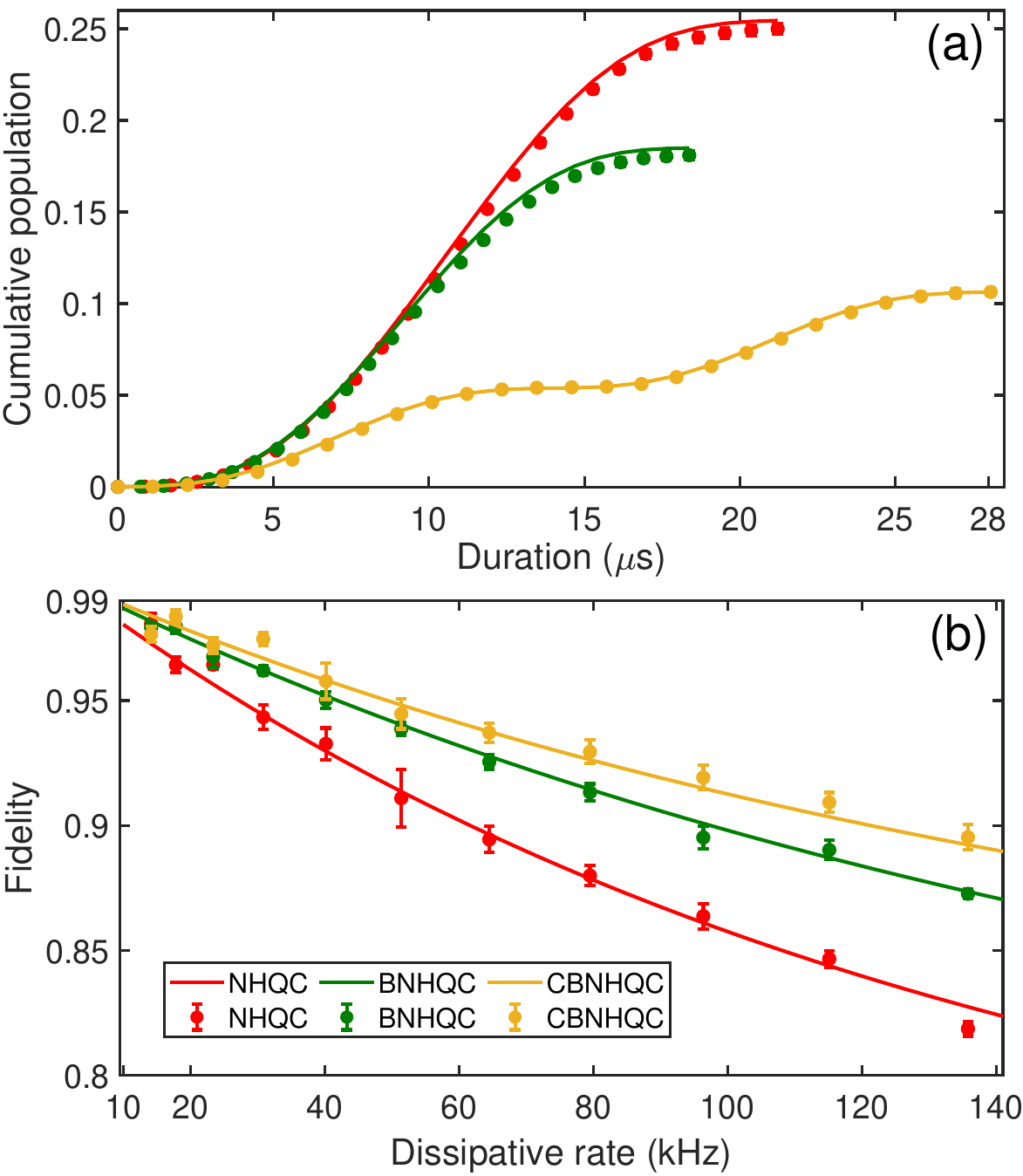}
\caption{(a) The accumulated population in the auxiliary excited state $\ket{a}$.  (b) The comparison of robustness against the dissipation rate $\kappa$ based on the experimental fidelities of the output states after undertaking the $\sqrt{X}$ gate. Here the parameters are selected same as Fig.\ref{fig1}.}
\label{fig:decay}
\end{figure}

To further characterize these protocols, we implement the quantum process tomography of gate~\cite{ref48,ref49} by preparing the system into the specific input state set $\{\ket{g}$, $\ket{e}$, $\ket{g}_x$,$\ket{g}_y\}$ and reconstructing the process matrix $\chi$ by measuring the corresponding output density matrices set. As shown in Fig.~\ref{fig:matrix}, the process fidelities ($\mathcal{F}=\sqrt{\text{Tr}[\chi_{\rm exp}\chi_{\rm id}]}$) for NHQC, BNHQC, and CBNHQC are 98.4(2)\%, 98.8(3)\%, and 99.5(2)\%, respectively, also consistent with the corresponding numerical calculation of 98.8\%, 99.2\%, and 99.4\%. Therefore, the CBNHQC shows an obvious fidelity advantage than the others. In Fig.~\ref{fig:decay}(a), it reveals the reason that the smaller accumulated population in auxiliary excited state $\ket{a}$ can effectively decrease the influence of decoherence on the evolution process of state.
 
\begin{figure}[t]
\includegraphics[width=8.3cm,height=9.3cm]{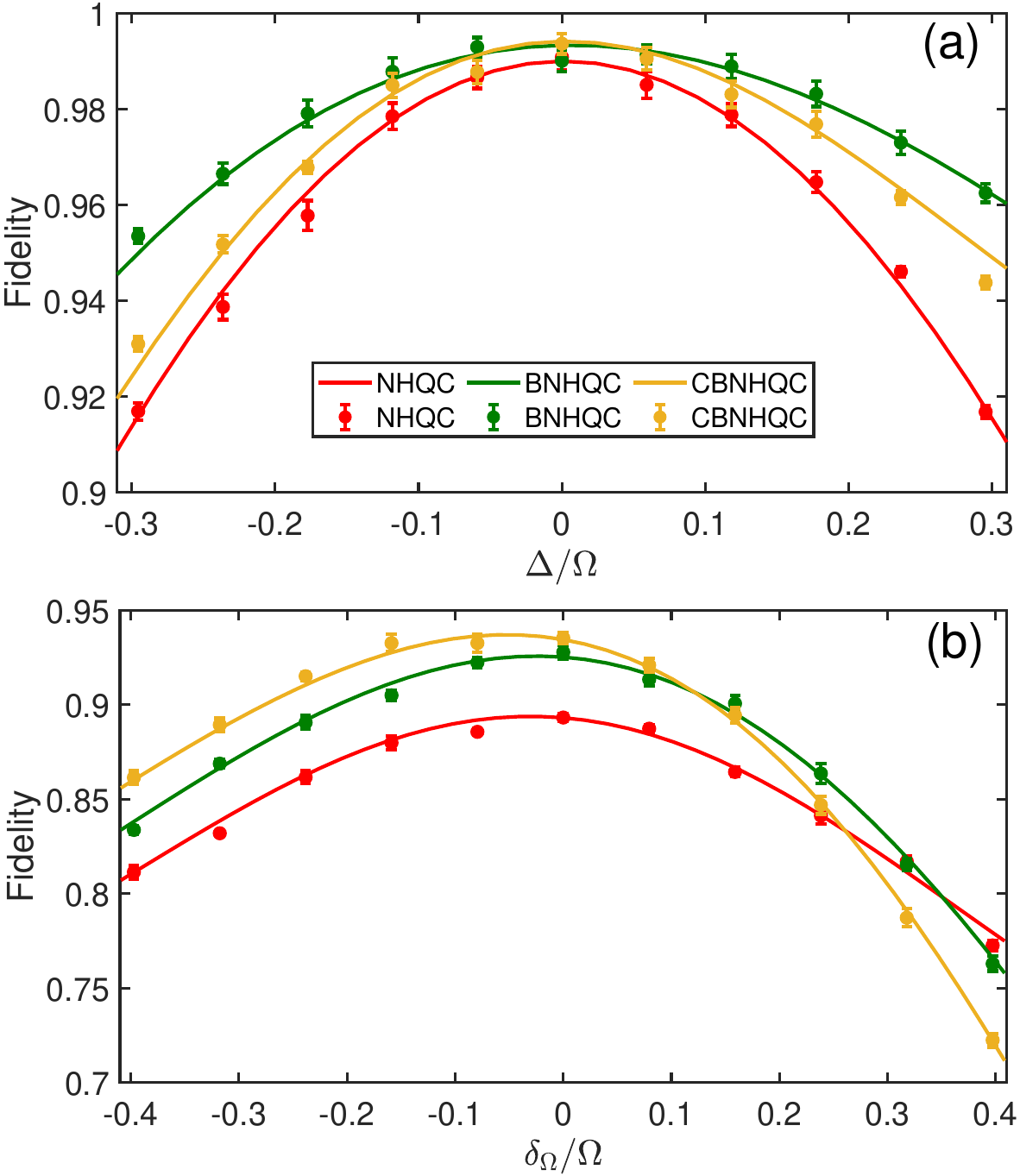}
\caption{Robustness of different schemes against the control errors, where (a) for resonance frequency error $\Delta$ of qubit and (b) for the Rabi frequency error $\delta_{\Omega}$ of the qubit. Here the Rabi frequency $\Omega/2\pi=33.3$~kHz and the dissipation rate $\kappa=66.7$~kHz.}
\label{fig:robust}
\end{figure}

To discuss the influence of decoherence, we further control the intensity of the 854~nm laser to increase the effective dissipation rate $\kappa$ of the $\ket{a}$ state. As shown in Fig.~\ref{fig:decay}(b), the experimental results are well consistent with the numerical simulation results and demonstrate the robustness of control protocols to dissipation. Besides, it also shows that the CBNHQC scheme maintains the highest fidelity, followed by the BNHQC scheme, while the NHQC scheme exhibits the lowest fidelity together with a significant deviation from CBNHQC and BNHQC. The reason is that compared with CBNHQC, the population of the excited state $\ket{a}$ is relatively high in the NHQC scheme, while the duration of NHQC is longer than BNHQC. Thus, the stronger interaction between the system and environment makes it more susceptible to dissipation and reduces the fidelity of the quantum gate.

 
Finally, we justify the advantages of the BNHQC by experiencing two control errors. To deliberate on this robustness, we have explored a wide range of the deviation from the correct values by using a larger decay rate and smaller Rabi frequency, and adjust the center frequency and Rabi frequency of the bichromatic laser to realize the control errors of detuning and Rabi frequency, respectively. As shown in Fig.~\ref{fig:robust}, BNHQC scheme exhibits the best robustness against detuning errors, while the CBNHQC scheme demonstrates the best robustness against coupling strength errors. The reason is that the BNHQC scheme has a shorter evolution time, which reduces the influence of detuning errors, while the CBNHQC scheme effectively suppresses the population of excited state, thereby reducing the influence of coupling strength errors. Besides, it also shows that the Rabi errors is more important than the detuning errors. NHQC also shows an approximately symmetrical robustness to the direction of errors, while BNHQC and CBNHQC demonstrate an asymmetrical robustness that they have a more robustness to the positive errors of detuning and negative errors of Rabi frequency.


In summary, we experimentally demonstrated a universal brachistochrone nonadiabatic holonomic quantum gate scheme in a trapped-ion system and used the $\sqrt{X}$ gate as a representative example to benchmark its performance. By comparing the fidelity of constructing this gate, we have compared the conventional NHQC, time-optimal BNHQC and composite CBNHQC under dissipation, detuning errors, and Rabi frequency errors. The results show that BNHQC shortens the gate duration relative to conventional NHQC, thus maintaining a higher fidelity and stronger robustness, while CBNHQC provides an improved suppression of certain systematic control errors and also shows a higher fidelity at the cost of a longer evolution time. We also show that the key point of keeping high fidelity and strong robustness is decreasing the accumulated population in the excited state in the building process of gate. These observations clarify the practical trade-off between speed and robustness in holonomic quantum control and support trapped-ion platforms as a promising setting for fast, resilient, and scalable geometric quantum gates.

\textit{Acknowledgement}$-$This work is supported by the National Key Research and Development Program of China under Grant No. 2022YFA1404500, by Cross-disciplinary Innovative Research Group Project of Henan Province under Grant No. 232300421004, National Natural Science Foundation of China under Grant Nos. 1232410, U24A2015, U21A20434, 12074346, 12274376, 12374466, 12074232, 12125406, 12504591, by Natural Science Foundation of Henan Province under Grant Nos. 232300421075, 242300421212, 252300421207, by Major science and technology project of Henan Province under Grant No. 221100210400, by the China Postdoctoral Science Foundation under Grants No. BX20250179.

\appendix

\end{document}